\documentclass[aps,prl,reprint,showpacs,byrevtex]{revtex4-2}
\let\oldhat\hat
\renewcommand{\hat}[1]{\oldhat{\mathbf{#1}}}

\usepackage{times,mathptm}
\usepackage[dvips]{graphicx,color}

\begin{document}
\title{Origin of Charge Density Wave in Layered Kagome Metal CsV$_3$Sb$_5$}
\author{Chongze Wang, Shuyuan Liu, Hyunsoo Jeon, and Jun-Hyung Cho$^{*}$}
\affiliation{Department of Physics, Research Institute for Natural Science, and Institute for High Pressure, Hanyang University, 222 Wangsimni-ro, Seongdong-Ku, Seoul 04763, Republic of Korea}
\date{\today}

\begin{abstract}
Using first-principles calculations, we identify the origin of the observed charge density wave (CDW) formation in a layered kagome metal CsV$_3$Sb$_5$. It is revealed that the structural distortion of kagome lattice forming the trimeric and hexameric V atoms is accompanied by the stabilization of quasimolecular states, which gives rise to the opening of CDW gaps for the V-derived multibands lying around the Fermi level. This Jahn-Teller-like instability having the local lattice distortion and its derived quasimolecular states is a driving force of the CDW order. Specifically, the saddle points of multiple Dirac bands near the Fermi level, located at the $M$ point, are hybridized to disappear along the $k_z$ direction, therefore not supporting the widely accepted Peierls-like electronic instability due to Fermi surface nesting. It is further demonstrated that applied hydrostatic pressure significantly reduces the interlayer spacing to destabilize the quasimolecular states, leading to a disappearance of the CDW phase at a pressure of ${\sim}$2 GPa. The presently proposed underlying mechanism of the CDW order in CsV$_3$Sb$_5$ can also be applicable to other isostructural kagome lattices such as KV$_3$Sb$_5$ and RbV$_3$Sb$_5$.
\end{abstract}
\maketitle


The two-dimensional (2D) kagome lattice composed of uniformly tiled triangles and hexagons [see Fig. 1(a)] possesses a unique electronic structure characterized by symmetry-enforced Dirac points and flat bands, thereby providing a novel platform to investigate various topological and correlated phenomena such as quantum anomalous Hall effect~\cite{AHC-Mn3Sn-Nat2015, AHC-Mn3Ge-Sci2016, AHC-Co3Sn2S2-Nat2018, AHC-Fe3Sn2-Nat2018}, charge fractionalization~\cite{Brien-PRB2010, Ruegg-PRB2011}, and spin liquid~\cite{SpinLiquid-Nat2010, SpinLiquid-Sci2011}. Recently, a new family of nonmagnetic layered kagome metals AV$_3$Sb$_5$ (A = K, Rb, and Cs)~\cite{AV3Sb5-PRM2019} has attracted enormous attention due to their intriguing electronic properties of nontrivial topological states~\cite{CsV3Sb5-Z2-PRL2020, KV3Sb5-AHC-AS2020, CsV3Sb5-AHC_SC-PRB2021, KV3Sb5-chrialCDW-Nat.Mat2021}, charge density wave (CDW)~\cite{KV3Sb5-chrialCDW-Nat.Mat2021, CsV3Sb5-2x2x2CDW-PRX, CsV3Sb5-CDW_SC-NC2021, CsV3Sb5-SC_CDW-PRL2021}, and superconductivity~\cite{CsV3Sb5-CDW_SC-NC2021, CsV3Sb5-SC_CDW-PRL2021, KV3Sb5_SC_Z2_PRM2021, RbV3Sb5-SC-CPL2021, CsV3Sb5_SC_PRB2021}. These kagome compounds have the Fermi level $E_F$ close to the saddle points of linearly dispersive Dirac bands, located at the $M$ point in the Brillouin zone. It was thus proposed that the van Hove singularities (vHs) arising from such saddle points could be attributed to the emergence of CDW order via Fermi surface (FS) nesting~\cite{Peiers, Kohn-PRL1959, KagomeFS-PRL2013, KagommeCDW-PRB2013}. However, the origin of the CDW formation in AV$_3$Sb$_5$ still remains elusive~\cite{AV3Sb5-B.Yan-PRL2021, AV3Sb5-PRB2021, CsV3Sb5-CDW_origin-PRB2021, AV3Sb5-B.Yan-PRL2021, KagomevHs-PRB2021, AV3Sb5-CDW-phon_nomaly, KV3Sb5-optical-CDW, CsV3Sb5-phonon}.

The CDW phases of AV$_3$Sb$_5$ have been experimentally observed at temperatures below about 80$-$100 K~\cite{KV3Sb5-chrialCDW-Nat.Mat2021, CsV3Sb5-SC_CDW-PRL2021, CsV3Sb5-CDW_SC-NC2021, CsV3Sb5-2x2x2CDW-PRX}. Using various experimental tools~\cite{KV3Sb5-chrialCDW-Nat.Mat2021, CsV3Sb5-2x2x2CDW-PRX, CsV3Sb5-SC_CDW-PRL2021, CsV3Sb5-SC_CDW-PRL2021, CsV3Sb5-CDW_SC-NC2021} and density-functional theory (DFT) calculations~\cite{AV3Sb5-B.Yan-PRL2021,CsV3Sb5-phonon}, such a CDW transition was revealed to form a three-dimensional (3D) 2${\times}$2${\times}$2 lattice reconstruction, in which the 2D kagome lattice exhibits a 2${\times}$2 structural distortion with three different V$-$V bond lengths [see Fig. 1(b)]. Hereafter, this CDW structure containing the two trimers and one hexamer of V atoms within a 2${\times}$2 unit cell is termed the tri-hexagonal (TrH) structure. Upon cooling down to temperatures below ${\sim}$3 K, the CDW phases were observed to exhibit a superconductivity~\cite{CsV3Sb5-SC_CDW-PRL2021, CsV3Sb5-CDW_SC-NC2021, KV3Sb5_SC_Z2_PRM2021, RbV3Sb5-SC-CPL2021, CsV3Sb5-AHC_SC-PRB2021}, indicating the coexistence of CDW and superconductivity. Based on the hypothesis that the AV$_3$Sb$_5$ compounds have the FS with the three nesting vectors $Q_1$ = (0.5, 0), $Q_2$ = (0, 0.5), $Q_3$ = (0.5, 0.5) in 2D reciprocal lattice units between neighboring vHs, it has been widely accepted that the inter-scattering of the electronic states connected by the three $Q_i$ vectors drives the 2${\times}$2 CDW transition~\cite{KagomeFS-PRL2013, AV3Sb5-B.Yan-PRL2021, KagommeCDW-PRB2013, CsV3Sb5-CDW_origin-PRB2021, AV3Sb5-PRB2021, KagomevHs-PRB2021}. In contrast to this FS nesting-driven Peierls-like electronic instability mechanism~\cite{Peiers, Kohn-PRL1959}, the CDW formation can also be driven by a momentum-dependent electron-phonon coupling (EPC) mechanism involving the concerted action of electronic and ionic subsystems~\cite{Marin-PRB2008, Barisic-PRL1970, 2H_NbSe2-PRL2011, Xuetao-PANS2015, Pouget-PRB2021}. Recent angle-resolved photoemission spectroscopy (ARPES) experiments observed the CDW gaps for the V-derived multibands at $E_F$ as well as away from $E_F$~\cite{KV3Sb5-CDW_Gap, RbV3Sb5-CDW_Gap, CsV3Sb5-CDW_Gap1, CsV3Sb5-CDW_Gap2}, suggesting that the EPC plays an essential role in forming the CDW phase in AV$_3$Sb$_5$. However, a combined inelastic x-ray scattering, Raman spectroscopy, and ARPES study~\cite{AV3Sb5-CDW-phon_nomaly} failed to observe acoustic phonon anomalies near the CDW wavevector $Q_i$ in AV$_3$Sb$_5$ and proposed another mechanism of many-body correlations and excitonic effects with particle-hole condensation.

In this Letter, we focus on CsV$_3$Sb$_5$ to identify the origin of CDW using first-principles DFT calculations. We reveal that the CDW instability with the structural transition to the 2D TrH structure is accompanied by the generation of the bonding and antibonding quasimolecular states in the trimeric and hexameric V atoms, leading to the opening of CDW gaps for the V-derived multibands around $E_F$. It is thus proposed that the major driving force of the CDW order is attributed to a Jahn-Teller effect~\cite{Hughes1977, Whangbo1992} possessing both the local structural distortion with the V trimer/hexamer and the formation of quasimolecular states~\cite{QMO-Mazin2012, Seho-TaS2}. Furthermore, we find that the FS shape of the pristine phase little changes with respect to pressure, but the CDW order is suppressed to disappear at a pressure of ${\sim}$2 GPa~\cite{CsV3Sb5-CDW_SC-NC2021, CsV3Sb5-SC_CDW-PRL2021}. Thus, the pressure-induced disappearance of the CDW phase is unlikely to be associated with FS nesting, but can be explained in terms of the destabilization of the quasimolecular states due to the significantly reduced interlayer spacing. Our findings are rather generic and hence, other isostructural kagome lattices such as KV$_3$Sb$_5$ and RbV$_3$Sb$_5$ can also adopt the same underlying mechanism for their CDW formations.

We first optimize the atomic structure of the pristine phase of CsV$_3$Sb$_5$ using the DFT calculation without including spin-orbit coupling (SOC)~\cite{method}. As shown in Fig. 1(a), the pristine phase crystallizes in the hexagonal space group $P6$/$mmm$ (No. 191) with the stacking of Cs hexagonal layer, upper Sb honeycomb layer, V$_3$Sb kagome layer containing a hexagonal Sb sublattice centered on each V hexagon, and lower Sb honeycomb layer. Here, the kagome layer has an identical nearest neighbor V$-$V bond length $d_{\rm V-V}$ = 2.726 {\AA}. Figure 2(a) shows the projected band structure of the pristine phase, together with the partial density of states (PDOS). We find that the V $d_{xz}$ and $d_{yz}$ orbitals are dominant components in the electronic states around $E_F$, compared to other orbitals such as the V $d_{xy}$, $d_{x^2-y^2}$, $d_{z^2}$, and Sb $p$ orbitals (see Fig. S1 in the Supplemental Material~\cite{SM}). In Fig. 2(b), we display the four FS sheets having different orbital characters: i.e., the circular sheet $FS_1$ composed of the Sb $p_z$ orbital, one hexagonal-shaped sheet $FS_2$ (at $k_z$ = 0) composed of the V $d_{xy}$, $d_{x^2-y^2}$, and $d_{z^2}$ orbitals, and two hexagonal-shaped sheets $FS_3$ and $FS_4$ composed of the V $d_{xz}$ and $d_{yz}$ orbitals. Here, $FS_2$ ($FS_3$) originates from the Dirac bands touching at the Dirac point located at $-$0.34 ($-$1.25) eV below $E_F$, while $FS_4$ is formed by one side of the Dirac bands touching at 0.85 eV above $E_F$ and another state $S_4$ [see Fig. 2(a)]. The existence of such four FS sheets and their shapes agree well with recent ARPES data~\cite{CsV3Sb5-ARPES-Kang, CsV3SB5-ARPES-Luo, CsV3Sb5-ARPES-Ho}.

\begin{figure}[h!t]
\includegraphics[width=8.5cm]{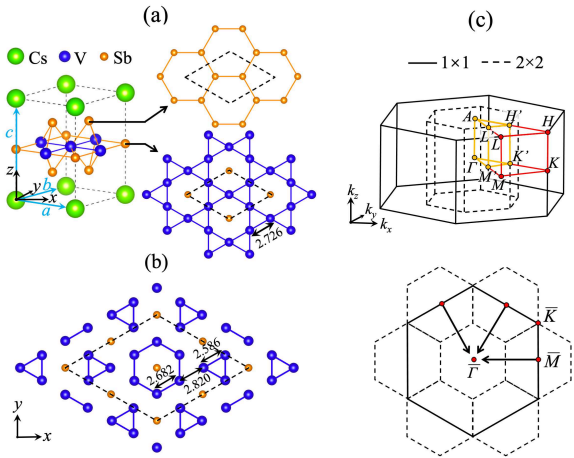}
\caption{ (a) Optimized structure of the pristine phase of CsV$_3$Sb$_5$, together with the top views of the Sb honeycomb and V$_3$Sb kagome layers. In (b), the top view of the V$_3$Sb kagome layer in the CDW phase is displayed. The numbers in (a) and (b) represent $d_{\rm V-V}$ in {\AA}. The Brillouin zones of the 1${\times}$1 pristine and 2${\times}$2 CDW phases are given in (c). On the bottom, the $\overline{M}$ points (along the $M-L$ line) are folded onto the $\overline{\Gamma}$ point (along the ${\Gamma}-A$ line) of the 2${\times}$2 Brillouin zone.}
\label{figure:1}
\end{figure}

\begin{figure}[h!t]
\includegraphics[width=8.5cm]{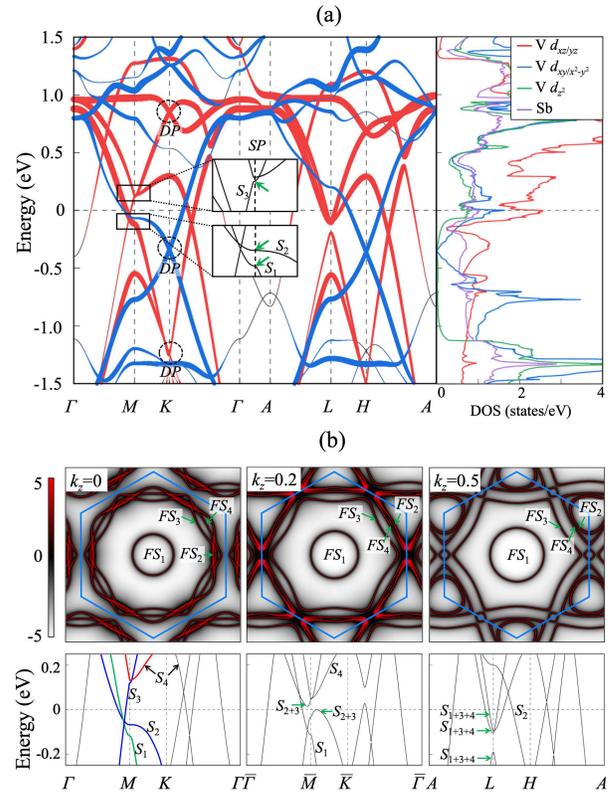}
\caption{ (a) Calculated band structure and PDOS of the pristine phase of CsV$_3$Sb$_5$. The bands are projected onto V $d_{xz}$/$d_{yz}$ (red) and $d_{xy}$/$d_{x^2-y^2}$ (blue) orbitals, where the radii of circles are proportional to the weights of the corresponding orbitals. In the inset, three saddle points ($SP$) at the $M$ point are indicated by arrows, and three Dirac points ($DP$) are marked by circles. In (b), four FS sheets~\cite{method} are displayed at $k_z$ = 0, 0.2, and 0.5, together with the corresponding band structures. The spectral function $A$($k$,$E_F$)~\cite{method} on the FS is drawn using the color scale.}
\label{figure:2}
\end{figure}

\begin{figure*}[h!t]
\includegraphics[width=17.0cm]{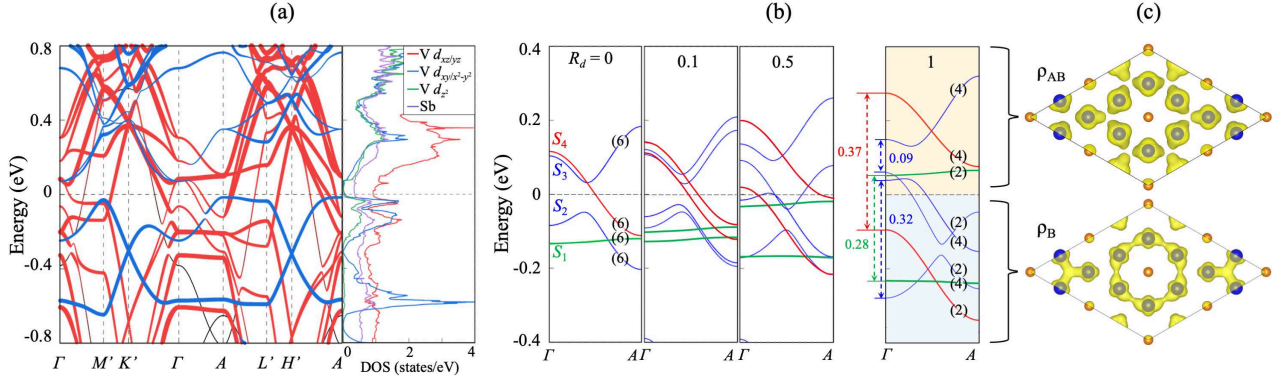}
\caption{ (a) Calculated band structures and PDOS of the 2$\times$2$\times$1 TrH CDW phase of CsV$_3$Sb$_5$. The PDOS unit is chosen as states/eV per 1$\times$1 unit cell. In (b), the band structures along the ${\Gamma}-A$ line are given as a function of $R_d$. The CDW gaps (in eV) of the $S_1$, $S_2$, $S_3$, and $S_4$ states are given in $R_d$ = 1. The numbers in parentheses represent degeneracy. In (c), the bonding (antibonding) charge density ${\rho}_{\rm B}$ (${\rho}_{\rm AB}$), integrated over the partially occupied (unoccupied) states, is displayed with an isosurface of 0.02 $e$/{\AA}$^3$.}
\label{figure:3}
\end{figure*}

According to previous studies of CsV$_3$Sb$_5$~\cite{KagomeFS-PRL2013, AV3Sb5-B.Yan-PRL2021, KagommeCDW-PRB2013, CsV3Sb5-CDW_origin-PRB2021, AV3Sb5-PRB2021, KagomevHs-PRB2021}, FS nesting between neighboring vHs formed by the saddle points of Dirac bands at the $M$ point was mostly believed to play a vital role in the formation of CDW order. Indeed, there are three saddle points associated with the $S_1$ $S_2$, and $S_3$ states, in the vicinity of $E_F$, i.e., at $-$0.14, $-$0.07, and 0.13 eV [see the insets in Fig. 2(a)]. However, these positions of saddle points do not coincide with the PDOS peaks located at $E_F$. In order to understand this incoincidence, we analyze the evolution of the four FS sheets with increasing $k_z$ and their band dispersions. As shown in Fig. 2(b), the saddle points of the $S_2$ and $S_3$ states shift toward $E_F$ to touch with each other around $k_z$ = 0.2. The resulting hybridization of the $S_2$ and $S_3$ states enhances their PDOS values at $E_F$. Note that the bright red color in the corresponding FSs, i.e., $FS_2$ and $FS_4$, indicates high DOS at $E_F$ [see Fig. 2(b)]. As $k_z$ approaches 0.5, $S_3$ and $S_4$ further shift downward to hybridize with $S_1$, leading to the disappearance of saddle points at the $\overline{M}$ point. It is noted that, as $k_z$ increases, $FS_2$ exhibits a dramatic change from a hexagonal shape at $k_z$ = 0 to a circular shape around the $\overline{K}$ [equivalent to $H$ in Fig. 1(c)] point at $k_z$ = 0.5 [see Fig. 2(b)]. This change in the topology of $FS_2$ around $k_z$ = 0.2 significantly influences the FS nesting property. In order to quantitatively assess the strength of FS nesting~\cite{Marin-PRB2008}, we calculate the low-frequency limit of the imaginary part of the electronic susceptibility, defined as ${\chi_0}^{''}$($q$) $=$ $\sum_{nm} \int d\mathbf{k} \delta(\varepsilon_{n\mathbf{k}}-E_{F}) \delta(\varepsilon_{m\mathbf{k+q}}-E_{F})$~\cite{Marin-PRB2008}. As shown in Fig. S2 in the Supplemental Material~\cite{SM}, ${\chi_0}^{''}$($q$) exhibits broad peaks around the $M$ point, indicating that the FS nesting-driven instability with a phonon softening at $Q_i$ is unlikely a driving force of the CDW formation in CsV$_3$Sb$_5$.

Next, we study the structural and electronic properties of the CDW phase having the TrH structure. Figure 1(b) displays the top view of the optimized TrH structure, showing three different bond lengths $d_{\rm V-V}$ = 2.586, 2.682, and 2.820 {\AA}. We find that the 2${\times}$2${\times}$1 TrH structure is more energetically favored than the pristine structure as well as the so-called Star of David CDW structure (see Fig. S3 in the Supplemental Material~\cite{SM}) by 14.13 and 5.38 meV per pristine unit cell, respectively, in good agreement with previous DFT calculations~\cite{AV3Sb5-B.Yan-PRL2021}. It is noted that the 2${\times}$2${\times}$2 TrH structure becomes more stable than the 2${\times}$2${\times}$1 one by ${\sim}$3.92 meV per pristine unit cell. Since the band dispersions of the two TrH structures along the ${\Gamma}-M'-K'-{\Gamma}$ line in the $k_x$-$k_y$ plane are similar to each other [see Figs. 3(a) and S4 in the Supplemental Material~\cite{SM}], we hereafter concentrate on the 2${\times}$2${\times}$1 TrH structure that preserves the key physics of the CDW formation in the 2D kagome lattice. In Fig. 3(a), we find that the 2${\times}$2${\times}$1 CDW phase suppresses the PDOS for all the constituent orbitals at $E_F$, compared to those of the pristine phase [see Fig. 2(a)]. Especially, PDOS values for the V $d_{xy}$, $d_{x^2-y^2}$, and $d_{z^2}$ orbitals become nearly negligible at $E_F$ with a gap of ${\sim}$0.06 eV. On the other hand, PDOS values for the V $d_{xz}$, $d_{yz}$, and Sb $p$ orbitals are finite at $E_F$, which can lead to the coexistence of CDW and superconductivity, as observed by several recent experiments~\cite{CsV3Sb5-SC_CDW-PRL2021, CsV3Sb5-CDW_SC-NC2021}.

To explore the relationship between the lattice distortion and CDW-induced gap ${\Delta}_{\rm CDW}$, we calculate the band structure as a function of $R_d$, defined as a ratio $d$/$d_0$ where $d$ represents the lateral shifted displacement of each V atom from the pristine toward the CDW phase ($d_0$ represents the full displacement). The calculated band dispersions along the ${\Gamma}-A$ line show that ${\Delta}_{\rm CDW}$ values of the $S_1$ and $S_4$ states increase with increasing $R_d$, reaching 0.28 and 0.37 eV at the ${\Gamma}$ point, respectively [see Fig. 3(b)]. Meanwhile, the $S_2$ and $S_3$ states hybridize with each other with increasing $R_d$, reaching the corresponding ${\Delta}_{\rm CDW}$ values of 0.32 and 0.09 eV, respectively. The existence of ${\Delta}_{\rm CDW}$ over the full range of $k$-space [e.g., see along the ${\Gamma}-A$ and $A-L'$ lines in Fig. 3(a)] is consistent with the ARPES data of CsV$_3$Sb$_5$~\cite{CsV3Sb5-CDW_Gap1, CsV3Sb5-CDW_Gap2} and KV$_3$Sb$_5$~\cite{KV3Sb5-CDW_Gap}, which exhibited the momentum-dependent CDW gaps for all the V-derived FS sheets. Therefore, the present results of ${\Delta}_{\rm CDW}$ do not support the previously proposed FS-nesting mechanism for the CDW formation in CsV$_3$Sb$_5$, where the $Q_i$-connected electronic states at $E_F$ are involved to generate ${\Delta}_{\rm CDW}$.

For $R_d$ = 0, the $S_1$, $S_2$, $S_3$, and $S_4$ states have the six-fold degeneracy along the ${\Gamma}-A$ line [see Fig. 3(b)], due to the band folding in the 2${\times}$2${\times}$1 CDW periodicity, i.e. the ${\Gamma}-A$ line is folded onto the three $M-L$ lines of the 1${\times}$1${\times}$1 Brillouin zone [see Fig. 1(c)]. As $R_d$ increases, the six-fold degeneracy is split into the two-fold and four-fold degeneracies~\cite{note1} [see Fig. 3(b)]. This band split reducing its degeneracy is induced by the structural reconstruction of V trimers and hexamers. In order to characterize such a CDW-induced electronic structure change, we plot the partial charge densities ${\rho}_{\rm B}$ and ${\rho}_{\rm AB}$ in Fig. 3(c), integrated over the occupied states between $E_F-$0.4 eV and $E_F$ and the unoccupied states between $E_F$ and $E_F$+0.4 eV, respectively. We find that the charge characters of ${\rho}_{\rm B}$ and ${\rho}_{\rm AB}$ represent the bonding and antibonding quasimolecular states originating from trimeric and hexameric V atoms, respectively: i.e., ${\rho}_{\rm B}$ exhibits a charge overlap connecting a trimer as well as a charge delocalization encircling a hexamer, whereas ${\rho}_{\rm AB}$ is evenly distributed over each V atom. These charge characters of the quasimolecular states are more distinct as $R_d$ increases (see Fig. S5 of the Supplemental Material~\cite{SM}). Consequently, such quasimolecular states having delocalized electrons contribute to lower the energy of the TrH CDW structure. Since the pristine structure is dynamically unstable with the presence of imaginary phonon frequencies around the $M$ and $L$ points and the total energy decreases monotonically with increasing $R_d$ (see Fig. S6 in the Supplemental Material~\cite{SM}), we can say that the 2${\times}$2 TrH CDW phase is significantly stabilized by the formation of the bonding and antibonding quasimolecular states with ${\Delta}_{\rm CDW}$. It is thus likely that the Jahn-Teller-like instability having the lattice distortion and its derived quasimolecular states is a driving force of the CDW formation in CsV$_3$Sb$_5$.

It is worth noting that each band in the pristine and CDW phases has twofold degenerate energy levels (i.e., the Kramers degeneracy) enforced by space-inversion and time-reversal symmetries. As shown in Fig. 2(a) and 3(a), there are many fourfold degenerate band crossings between the V $d_{xz}$/$d_{yz}$ and $d_{xy}$/$d_{x^2-y^2}$ orbitals on $k_z$ = 0 and 0.5 because of the opposite parity eigenvalues of the mirror symmetry $M_z$ about the $x$-$y$ plane~\cite{symmetry_CPB2016}. These band crossings are lifted through the inclusion of SOC~\cite{symmetry_CPB2016}. By analyzing the parity of the wave function at the time-reversal invariant momentum points, we confirm that both the pristine and CDW phases possess the same $Z_2$-type band topology (See Fig. S7 in the Supplemental Material~\cite{SM}), consistent with previous theoretical calculations~\cite{AV3Sb5-B.Yan-PRL2021,CsV3Sb5-Z2-PRL2020}. Given the presence of chiral charge order at the surface of an isostructural kagome material KV$_3$Sb$_5$~\cite{KV3Sb5-chrialCDW-Nat.Mat2021}, the present $Z_2$-type topological band structure of CsV$_3$Sb$_5$ can be enabled to produce a large anomalous Hall effect possibly due to a chirality-driven time-reversal symmetry breaking, as measured by transport experiments~\cite{CsV3Sb5-AHC_SC-PRB2021}.

\begin{figure}[h!t]
\includegraphics[width=8.5cm]{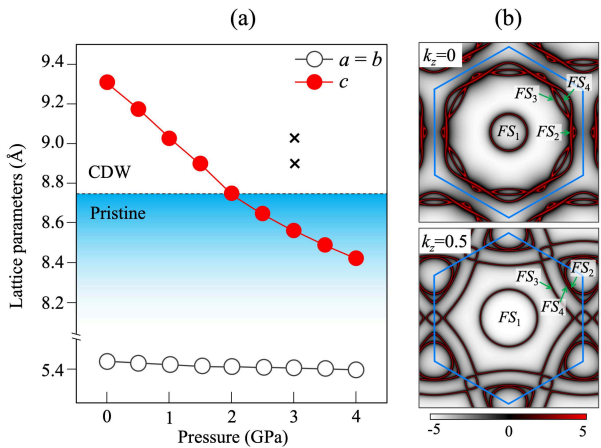}
\caption{(a) Calculated lattice parameters of CsV$_3$Sb$_5$ as a function of pressure and (b) four FS sheets at $k_z$ = 0 and 0.5, obtained at 2 GPa. In (a), the phase transition from the CDW to the pristine phase at 2 GPa is illustrated. The marks (${\times}$) represent the TrH CDW structures with two different $c$ values at 3 GPa.}
\label{figure:4}
\end{figure}

We further examine how the CDW instability changes under applied pressure. It is found that the CDW order is suppressed with increasing pressure, and finally disappears at a pressure of ${\sim}$2 GPa [see Fig. 4(a)], in good agreement with experimental data~\cite{CsV3Sb5-SC_CDW-PRL2021, CsV3Sb5-CDW_SC-NC2021}. Here, the calculated lattice parameters versus pressure shows that $a$ and $b$ values little decrease but $c$ value sharply decreases as pressure increases. This implies that the compressed TrH CDW phase could be much influenced by the $c$-axis lattice parameter. We find that at 3 GPa, the pristine phase is transformed into the TrH CDW phase with increasing $c$ value: i.e., the latter phase having the quasimolecular states becomes more stable than the former one by 6.95 and 13.67 meV per pristine unit cell at $c$ = 8.857 and 8.983 {\AA} [see the data marked (${\times}$) in Fig. 4(a)], respectively. Thus, the pressure-induced disappearance of the CDW phase at 3 GPa is likely attributed to the destabilization of the quasimolecular states with an increased interlayer interaction due to the significantly reduced $c$ value. It is, however, noted that the FS shape of the pristine structure little changes at 2 GPa [see Figs. 4(b)], manifesting that the FS nesting mechanism is unlikely associated with the pressure-induced disappearance of CDW.

To conclude, based on first-principles DFT calculations, we have identified that the CDW formation in CsV$_3$Sb$_5$ is driven by a Jahn-Teller-like instability having the local lattice distortion and its derived quasimolecular states, rather than the widely accepted Peierls-like electronic instability due to Fermi surface nesting. Since other isostructural kagome lattices such as KV$_3$Sb$_5$ and RbV$_3$Sb$_5$ have a similar Jahn-Teller effect as that of CsV$_3$Sb$_5$ (see Fig. S8 in the Supplemental Material~\cite{SM}), our findings shed light on the understanding of the origin of the CDW formation in a new family of nonmagnetic layered kagome metals AV$_3$Sb$_5$.

\vspace{0.4cm}




\noindent $^{*}$ Corresponding author: chojh@hanyang.ac.kr

\end{document}